\documentclass[aps,prd,twocolumn,groupedaddress]{revtex4}

\usepackage{graphics}
\usepackage{graphicx}
\usepackage{dcolumn}
\usepackage{bm}
\usepackage{epsfig}

\begin{document}


\newcommand{\be}{\begin{equation}}
\newcommand{\ee}{\end{equation}}
\newcommand{\bea}{\begin{eqnarray}}
\newcommand{\eea}{\end{eqnarray}}



\title{Analytic Gravitational-Force Calculations
 for Models of the Kuiper Belt,\\
with Application to the Pioneer Anomaly}

\author{Michael Martin Nieto}
\email{mmn@lanl.gov}
\affiliation{Theoretical Division, MS-B285, 
Los Alamos National Laboratory,
University of California, Los Alamos, New Mexico 87545}

\date{\today}


\begin{abstract}
We use analytic techniques to study the gravitational force that would be produced by different Kuiper-Belt mass distributions.  In particular, we study the 3-dimensional rings (and wedge) whose densities vary as the inverse of the distance, as a constant, as the inverse-squared of the distance, as well as that which varies according to the Boss-Peale model.  These analytic calculations yield physical insight into the physics of the problem.  They also verify that physically viable models of this type can produce neither the magnitude nor the constancy of the Pioneer anomaly.

\end{abstract}


\maketitle


\section{\label{intro}  Introduction}     

There has long been interest in the gravitational force that could be produced by the Kuiper Belt \cite{boss}.  It has been observed that total masses of much more than an Earth mass, $M_\oplus$, would lead to conflicts with orbital observations. 
(See, e.g., Refs. \cite{boss,KBjda} and Sec. VII-E of \cite{pioprd}.)  
Further, it has also been calculated that a Kuiper-Belt ring with a mass of this magnitude could not explain an acceleration the size of the Pioneer anomaly \cite{boss,KBjda,pioprd}.  This anomaly \cite{pioprd,pioprl} is the apparent unmodeled constant acceleration of the Pioneer spacecraft, observed between $\sim ~ 20-70$ Astronomical Units (AU), of magnitude
\be
a_P(20 \mathrm{AU} < r< 70 \mathrm{AU}) = - 
(8.74 \pm 1.33) \times 10^{-8} ~\mathrm{cm/s}^2 
\ee
which is directed approximately towards the Sun.

Even so, this type of Kuiper-Belt mechanism has remained a fascinating one as a possible explanation of the anomaly.  In particular, 
it has recently been proposed \cite{diego} that gravitation from the Kuiper Belt, modeled by a cylindrically-symmetric ring of matter whose density goes as 
\be
\rho_1(p) = \frac{\rho_1 ~L}{p}, ~~~~~ p = \sqrt{x^2 +y^2}\label{1/p},
\label{1/r-pot}
\ee
where 
\be
\rho_1= 1.74 \times 10^{-16} \mathrm{g/cm}^3, ~~~~~~~L = 20 ~\mathrm{AU},
\label{dparam}
\ee
can explain the {\it constant} anomaly. 
The ring has a width 
\be
R_1 =20~\mathrm{AU} ~\le~p~\le~100~\mathrm{AU} =R_2 
\ee 
and a thickness 
\be
2D = 2 ~\mathrm{AU}.  
\ee
The mass is thus 
\be
{\cal M}_{ring} = 4\pi\rho_1L~D (R_2-R_1)
            = 1.17 \times 10^{28}~\mathrm{g} = 1.96 M_\oplus.
\label{mring}
\ee

This proposal is somewhat surprising, given the observations noted above.  However, one is thereby motivated to take a different looks at the problem \cite{orfeu}.  Here we do so emphasizing analytic calculations.  This will help to better understand the underlying physics of the situation. 

To start, although it is well-known that a spherically symmetric ball with a density that goes as $1/r$ can produce a constant acceleration within the ball,  there {\it only} is a constant acceleration from a complete spherical ball, not from a shell.  Therefore, as we emphasize in the next section, with only a cylinder ring, not even a cylindrical disk, satisfying a constant acceleration is doubly hard to do.  Specifically, it can not come from an exact cylindrically-symmetric $1/p$ density.  
Indeed, although the appeal to Gauss' Law in Eq. (3) of \cite{diego} is correct, the argument that Eq. (4) of \cite{diego} implies there will be a constant acceleration within the ring is not exact. We will demonstrate this by   
specific analytical calculation.   

Before continuing, we note again that the mass of the model belt of Ref. \cite{diego} appears to be somewhat high, as has been determined elsewhere \cite{boss,KBjda,pioprd}.  Further, it is known that the amount of dust is {\it much} smaller than this, and the gravitational mass of the Kuiper Belt is dominated by large rocks and ices.  The interplanetary dust is actually supplied by collisions between the rocks and ices and lives for only of order 100,000 years in the inner solar system, an order of magnitude longer in the outer solar system.  Further, the dominant mass of the rocks and ices is overwhelming subject to gravity and not other forces.  Hence, there tend to be resonant concentrations in it vs. a smooth distribution \cite{mann1}-\cite{drag}.  

In Section \ref{sphere} we will describe the gravity of spherical balls and shells.  This is followed by an introduction to the gravity of cylindrically symmetric disks and rings in Section \ref{rings}.  (These objects are examined in both the complete 3-dimensional framework and also in the ``thin-ring" approximation, where the distribution in the $z$ direction is a $\delta$-function.)
In Section \ref{thinrings} we apply the ``thin-ring" approximation to 
both the $1/p$ model and the Boss-Peale model \cite{boss}.   We then go on to full 3-dimensional calculations.  In Sections \ref{3-d-1-r},  \ref{constring}, and \ref{sphwedge} we discuss, respectively, the $1/p$-density cylindrical ring, the constant-density cylindrical ring, and the $1/r^2$-density wedge (as well as the $1/p^2$-density ``thin ring").  We end with a discussion where we compare the results.  In particular, we compare the accelerations produced by the 3-dimensional $1/p$-, $1/r^2$-, and constant-density rings, as well as those from the Boss-Peale and $1/p^2$ ``thin-rings."   

We find, as expected, that neither the magnitude nor the shape of the Pioneer anomaly can be reproduced.  (For comparison, in our numerical plots we will adhere as much as possible to the model parameters of Eq. (\ref{dparam}).  However, since the basic formulae are analytic, they can be renormalized at will.)


\section{\label{sphere} Spherical balls and shells}

The $1/r^2$ gravitational force law yields that any spherically symmetric distribution with total mass $\mathcal{M}$ exerts a force outside that distribution that is proportional to the total mass divided by the square of the distance to the center of symmetry: $-G\mathcal{M}/r^2$.   Contrarily, if the observation point is inside a spherical distribution of mass, no force is exerted. 

This is an important result for understanding the effects of a general spherically symmetric density distribution, $\rho(r)$.   Since we are heading towards the $1/r$ distribution, consider density distributions
that go as 
\be
\rho(r) \rightarrow \frac{\rho_n(r) ~L^n}{r^n}, ~~~~
           -\infty \le n \le \infty. \label{1/rton}
\ee
Here $\{\rho_n,L\}$ give the overall normalizations in terms of some density and length scale.
These types of densities have long been studied by geophysicists.  They often like to think in terms of spherical distributions and shells of the Earth having different functional dependences and thus causing different gravity signals  
\cite{dzi81,geo}. 
{\it But note:} 
We are talking about spherical shells, {\it not} cylindrical rings. 

In the present study, we will concentrate on the distributions for 
$n=\{0,1,2\}$, the constant, $1/r$, and $1/r^2$ distributions. 
Specifically, start with the $1/r$ distribution, 
\be
\rho_1(r) = \frac{\rho_1 ~L}{r}.  \label{1/r}
\ee
It has a total mass out to a radius $R$ of 
\be
{\cal M}_{ball}(R) = 2\pi\rho_1~L~R^2.
\ee
(Of course, if the density distribution went to infinity there would be infinite mass.)  From the spherical symmetry condition mentioned before, we have that interior and exterior to the sphere  
\bea 
a_{ball}(r<R) &=& \frac{-G~{\cal M}_{ball}(r)}{r^2} 
                  = -G~2\pi \rho_1~L, \label{ball} \\
a_{ball}(r>R) &=&\frac{-G~{\cal M}_{ball}(R)}{r^2} 
                = \frac{-G ~2\pi\rho_1~L~R^2}{r^2}. \nonumber \\
&~&
\eea
That is, there is a constant acceleration inside the ball and the ordinary Newtonian inverse-square force outside the ball.  Even so, there remains a singularity at the origin since there the acceleration is a non-zero constant pointing radially in from all directions. 

If we now use the parameters of Ref. \cite{diego} given in Eq. (\ref{dparam}) above, $\rho_1= 1.74 \times 10^{-16}$ g/cm$^3$ and $L = 20$ AU,
then even the spherical ball of Eq. (\ref{ball}) would only produce an acceleration of magnitude
\bea 
a_{ball}(r<R) &=& -C_{ball} = -(2\pi G~\rho_1L) \nonumber \\
&=& -2.18 \times 10^{-8} ~\mathrm{cm/s}^2.
\eea
But this is smaller than $a_P$!  So, if an entire ball of this density can not cause the Pioneer anomaly, how can a disk, let alone a ring?

To continue, what if this were only a spherical shell (from $R_1= 20$ AU to $R_2=100$ AU)?  Then, even inside the shell the acceleration would not be constant.  By subtracting out the gravitational attraction of the mass interior to radius $R_1$ the acceleration is 
\bea
a_{shell}(0<r<R_1) &=& 0,  \\
a_{shell}(R_1<r<R_2)&=&-G~ 2\pi \rho_1L  
               + \frac{G~{\cal M}_{ball}(R_1)}{r^2}     \nonumber \\
          &=&-G~ 2\pi \rho_1L  + \frac{G~2\pi \rho_1~L~R_1^2}{r^2}, 
 \nonumber \\
&~&\\
a_{shell}(r>R_2) &=& \frac{-G~2\pi \rho_1~L~(R_2^2- R_1^2)}{r^2},
\eea
where we write
\bea
a_{shell}(r) &\equiv& -(2\pi G~\rho_1L)~g_{shell}(r) = -C_{ball}~g_{shell}(r), 
 \nonumber \\
&~&\\
{\cal M}_{shell} &=& 2 \pi \rho_1 L (R_2^2 - R_1^2) = 60 {\cal M}_{ring} 
 \nonumber \\
      &=& 7.03 \times 10^{29}~ \mathrm{g}. 
\eea

\begin{figure}[h]
 \begin{center}
\noindent    
\psfig{figure=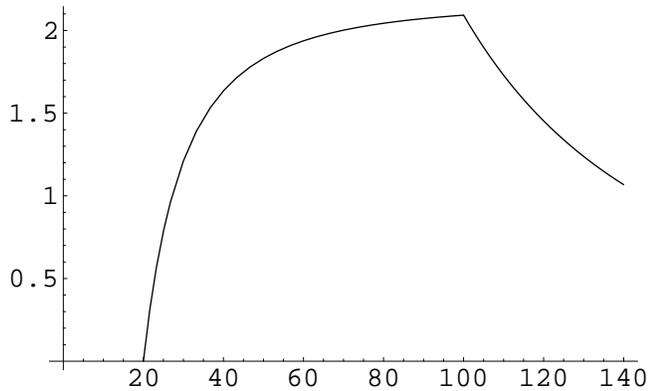,width=3.375in}
\end{center}
  \caption{A plot of $-a_{shell}(r)$  in units of $10^{-8}$ cm/s$^2$ vs. $r$ in AU. 
 \label{gshell}}
\end{figure}

Therefore, there is a constant acceleration towards the center of a 
spherical $1/r$-density distribution of matter given by Eq. (\ref{1/r}) {\it only} if the mass distribution goes all the way into the origin; that is, if it is a spherical ball, not a spherical shell.  In Figure \ref{gshell} we show $-a_{shell}(r)$ vs $r$ for the values $\{R_1,R_2\} = \{20,100\}$ AU.  This figure will be useful for comparison when we go to rings. 


Particular values of the acceleration are 
\be
-10^8~a_{shell}(\{10,60,120\}\mathrm{AU}) = 
           \{0, 1.94,1.45\}~\mathrm{cm/s}^2.
\label{a-shell}
\ee
However, even here with only the first 20 AU of the 100 AU ball deleted, the acceleration varies by an order 10\% in the outer half of the shell and rapidly decreases to zero interior to that.


\section{\label{rings} Cylindrical disks and rings}


\subsection{\label{full3-dring} Full 3-d disks and rings}

Now we go on to disks and rings.  We use a method inspired by techniques to   analyze \cite{cylinder} cylindrically-symmetric objects in laboratory big-G experiments \cite{heyl}-\cite{newman}. 
The general potential functional and acceleration  from a cylindrical symmetric ring are
\bea
{\cal V}(r) &=& V(r)/m_{test}, \\
{\cal V}(r)&=&-G\int^{+D}_{-D} dz \int^{R_2}_{R_1}{dp~p}~\rho(p)~\times
 \nonumber \\
&~&
\int^{2\pi}_0 \frac{d\phi}{\sqrt{p^2 +r_x^2 - 2r_xp\cos\phi +(r_z-z)^2}},
\label{3-dV} 
 \nonumber \\
&~&\\
a_x(r) &=&-G\frac{d}{dr_x} \left[{\cal V}(r)\right],  
\label{3-da} \\
a_z(r) &=&-G\frac{d}{dr_z} \left[{\cal V}(r)\right]. \label{3-dz}
\eea

In the above, by convention we take the component of the direction to the test mass in the plane of the ecliptic to be along the $x$ axis: $r_{\{x,y\}}\rightarrow r_x$. This is useful since we will concentrate on the case of axial symmetry.   We also observe that 
the $z$-component of the acceleration in Eq. (\ref{3-dz}), for general positions out of the ecliptic, is easier to handle \cite{lass,BT} in the ``thin-ring" approximation of the next subsection.  

We denote these various choices by:  
\be
{\mathbf{r}}  \rightarrow (r_x, ~0, ~r_z) ~
{\rightarrow}_{ecliptic} ~(r,~0,~0). 
\ee
(Note for future reference that, with cylindrical symmetry, 
the volume element, $p$, cancels the $(1/p)$ of a $\rho_1(p)$ density function.)  


\subsection{``Thin-ring" approximation}


\subsubsection{\label{genTR} General thin rings}

As an initial step, we start in the next section by using an analytic approximation,
\be
\rho(r) \rightarrow   2D\delta(z)~\rho(p).  
\ee
We can do this because $z$ is generally small compared to $p$ so the change in the overall result should be small and still symmetric about the z axis.  

This yields 
\bea
{\cal V}_{thin}(r)&=&-2GD \int^{R_2}_{R_1}{dp~p}~\rho(p) ~\times 
 \nonumber \\
&~&
\int^{2\pi}_0 \frac{d\phi}{\sqrt{p^2 +r_x^2 - 2r_xp\cos\phi + r_z^2}},
\label{2-dV} \\
a_{thin}(r) &=&2GD\frac{d}{dr_x}\left[
\int^{R_2}_{R_1}{dp~p}~\rho(p) \right. ~\times
 \nonumber \\
&~& \left.
\int^{2\pi}_0 \frac{d\phi}{\sqrt{p^2 +r_x^2 - 2r_xp\cos\phi + r_z^2}}
\right].  
\label{2-da}
\eea


\subsubsection{\label{numTR} Taking the $\mathbf{r}$-derivative first}

One tack that can be taken (and will be in Sections \ref{analytic-1/p} 
and \ref{boss-1/p} below) is to first perform the $r_x$-derivative in Eq. (\ref{2-da}), 
\bea 
a_{thin}(r) &=& -4GD\int^{R_2}_{R_1}{dp~p}~\rho(p) ~\times
 \nonumber \\
&~&
\int^{\pi}_0 \frac{d\phi~(r_x-p\cos\phi)}{[p^2 +r_x^2 - 2r_xp\cos\phi + r_z^2]^{3/2}}, 
\eea
and then do the $\phi$-integral.  
Going to the plane of the ecliptic, $r_z \rightarrow 0$, the result is 
\begin{widetext}
\bea
a_{thin}(r) &=& -4GD \int^{R_2}_{R_1}{dp~p}~\rho(p) ~
\left[\frac
{\mathbf{K}\left(\sqrt{\frac{-4pr}{r^2-2pr+p^2}}\right)}
{r\sqrt{r^2-2pr+p^2}} 
+\frac{(r-p)
\mathbf{E}\left(\sqrt{\frac{-4pr}{r^2-2pr+p^2}}\right)}
{r(r+p)\sqrt{r^2-2pr+p^2}}\right]    \label{KEints}\\
&=& -4GD \int^{R_2}_{R_1}{dp~p}~\rho(p) ~
\left[\frac
{\mathbf{K}\left(\sqrt{\frac{4pr}{r^2+2pr+p^2}}\right)}
{r(r+p)} 
+\frac{
\mathbf{E}\left(\sqrt{\frac{4pr}{r^2+2pr+p^2}}\right)}
{r(r-p)}\right], 
 \label{bossform}
\eea
where the last two equalities are related by 8.127 of Ref. \cite{GR} and 
the complete elliptic integrals of the first and second kind  (see 8.113 and 8.114 in \cite{GR}) are
\bea
\mathbf{K}(t) &\equiv& K(t^2) 
=\frac{\pi}{2}F\left(\frac{1}{2},\frac{1}{2};1;t^2\right) 
=\frac{\pi}{2}\left(1 + \frac{t^2}{4} + \frac{9~t^4}{64} + \dots +
\left[\frac{(2n-1)!!}{2^n n!}\right]^2t^{2n} + \dots \right),
\label{K}  \\
\mathbf{E}(t) &\equiv& E(t^2) 
=\frac{\pi}{2}F\left(-\frac{1}{2},\frac{1}{2};1;t^2\right) 
= \frac{\pi}{2}\left(1 - \frac{t^2}{4} - \frac{3~t^4}{64} - \dots -
\left[\frac{(2n-1)!!}{2^n n!}\right]\frac{t^{2n}}{2n-1} - \dots \right).
\label{E} 
\eea
\end{widetext}

This yields a physically intuitive $p$-integration that can be handled numerically \cite{boss}.   
We will use Eq. (\ref{bossform}) in Sections \ref{boss-1/p} and \ref{boss-peale} below. 


\section{\label{thinrings} Specific thin rings}


\subsection{\label{analytic-1/p}
Analytic, thin-ring, $\mathbf{1/p}$-density model}

Returning to Section \ref{genTR}, it turns out that the thin-ring, ${1/p}$-density problem is analytically solvable. 
If one does the $\phi$-integral before the r-differentiation in Eq. (\ref{2-da})
one can also do the second integral.   (Again note, for this $1/p$-density case, the density function cancels the $p$ in the volume element, making the integrals simpler.)    
Proceeding, the potential functional in the plane of the ecliptic is
\be
{\cal V}_{T/p}(r)=-G\rho_1~2DL\int^{R_2}_{R_1}dp
\int^{2\pi}_0 \frac{d\phi}{\sqrt{p^2 +r^2 - 2rp\cos\phi}}. \label{V(r)}
\ee
The $\phi$ integral is analytic and is (3.674.1 in \cite{GR})
\bea
I_\phi(r>p) &=&  \frac{4}{r} \mathbf{K}\left({p}\over{r}\right), \label{I1} \\
I_\phi(r<p) &=&  \frac{4}{p} \mathbf{K}\left({r}\over{p}\right). \label{I2}
\eea  
(Eqs. (\ref{V(r)}) and (\ref{I1}) demonstrate that for very large $r$ the potential goes to $-GM_{ring}/r$, as it should.)  

This means that the potentials outside, within, and inside of the ring are  
\bea
&~&{\cal V}_{T/p}(R_2<r) = -8G\rho_1L~D
\int^{R_2}_{R_1} \frac{dp}{r}\mathbf{K}\left({p}\over{r}\right), \\ 
&~&{\cal V}_{T/p}(R_1<r<R_2) = -8G\rho_1L~D\left[
\int^{r}_{R_1} \frac{dp}{r}\mathbf{K}\left({p}\over{r}\right) \right.
\nonumber \\ 
&~&~~~~~~~~~~~~~~~~~~~~~~~~~~~~~~+ \left.
\int^{R_2}_{r} \frac{dp}{p}\mathbf{K}\left({r}\over{p}\right) \right], \\
&~&{\cal V}_{T/p}(r<R_1) = -8G\rho_1L~D
\int^{R_2}_{R_1} \frac{dp}{p}\mathbf{K}\left({r}\over{p}\right) .
\eea
Changing variables to $t=p/r$ or $r/p$, respectively, and  
using the properties of the complete elliptic integral, the acceleration 
($a_{T/p} = -d{\cal V}_{T/p}/dr$) is 
\bea
&&a_{T/p}(r) = -C_1 ~ g_{T/p}(r),  \label{firsta} \\
&& C_1 =  8G\rho_1L = (4/\pi)~C_{ball} 
   = 2.779 \times 10^{-8} \mathrm{cm/s}^2, \label{C}\\
&& g_{T/p}(R_2<r) = 
\frac{DR_2}{r^2}\mathbf{K}\left(\frac{R_2}{r}\right) 
- \frac{DR_1}{r^2} \mathbf{K}\left(\frac{R_1}{r}\right),
\label{g1}\\
&& g_{T/p}(R_1< r <R_2) = \frac{D}{r}\mathbf{K}\left(\frac{r}{R_2}\right) 
- \frac{DR_1}{r^2}\mathbf{K}\left(\frac{R_1}{r}\right), 
\nonumber \\ &&
\label{g2}\\
&&g_{T/p}(r <R_1) = \frac{D}{r}\mathbf{K}\left(\frac{r}{R_2}\right) 
- \frac{D}{r}\mathbf{K}\left(\frac{r}{R_1}\right).
\label{g3}
\eea
This acceleration is  {\it not} a constant for $(R_1 < r < R_2)$.


\begin{figure}[h]
 \begin{center}
\noindent    
\psfig{figure=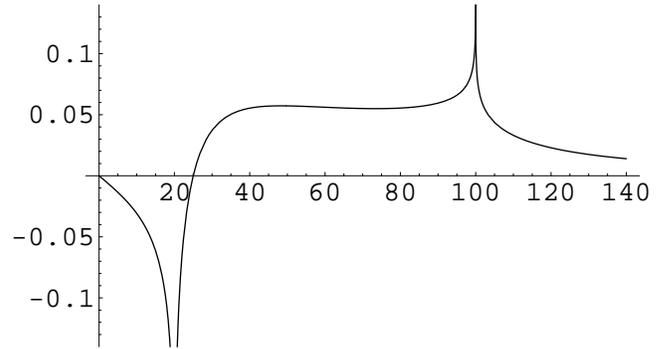,width=3.375in}
\end{center}
  \caption{A plot of $-a_{T/p}(r)$ in units of $10^{-8}$ cm/s$^2$ vs. $r$ in AU. 
 \label{2-dKB}}
\end{figure} 


Putting the remaining distances in terms of AU, in Figure \ref{2-dKB} we plot $-a_{T/p}(r)$  vs. $r$ using the parameters of Ref. \cite{diego}.   
One can note the general features.  Most importantly, the size of the acceleration within this model of the Kuiper Belt is about a factor of 100 smaller than the anomaly! 
In particular, specific values of the acceleration are 
\bea
&~& -a_{T/p}(\{10,60,120\}\mathrm{AU}) =
\nonumber \\ &~& 
           \{-0.0309, +0.0610,+0.0338\}\times 10^{-8}~\mathrm{cm/s}^2,
\label{a-T/p}
\eea
which can be compared to the values from a shell given in Eq. (\ref{a-shell}). The acceleration within the ring is of order 40 times smaller than that within the shell. 

Observe that $a_{T/p}(r)$ manifestly has other appropriate physical properties. First, 
$a_{T/p}(R_2 \ll r) \rightarrow -G{\cal M}_{ring}/r^2$.  Next, as it should on physical grounds,
$-a_{T/p}(r\rightarrow 0)\rightarrow 0_-$.   Analytically, 
Eqs. (\ref{K}), (\ref{firsta}), and (\ref{g3})  show that $-a_{T/p}(r)$ is slightly negative as $r\rightarrow 0$ and goes to zero in the limit.  

One also sees the breakdowns at 
$r=\{R_2,R_1\}$ where the $\mathbf{F}$ are singular because the arguments are unity.   (Here and later we will cut off the heights of the 2-d spikes.)  As we will see, these singularities result from having only a 2-d approximation for the non-smooth (hard-edged) ring.  When the density is continuous in the $p$ variable the spike singularity in the acceleration disappears, even for 2-d problems.  When the problem is 3-d, the spikes become finite cusps.  (See Section \ref{3-d-1-r}.)  

As observed, far out $a_{T/p}(r)$ goes as $1/r^2$.  As one comes in, approaches, and then passes $r=R_2$, the quantity $-a_{T/p}(r)$ starts to decrease since less mass is interior to the test point.  Within the interior of the ring, for a short distance  $a_{T/p}(r)$ is ``roughly," but not exactly, flat.  (It will be less constant in the true 3-d calculation.)  Further, as one gets closer to $R_1$ the acceleration changes sign because more mass begins pulling out rather than in.  
As predicted one sees that  $-a_{T/p}(r)$ is slightly negative as $r\rightarrow 0$  and it goes to zero at the origin.  


\subsection{\label{boss-1/p}
Another thin-ring, $\mathbf{1/p}$-density calculation}

We demonstrate here that an equivalent result for the ${1/p}$-density can be obtained by the method of Section \ref{numTR}.  This demonstration illuminates this method which will be useful in the following subsection.
 
If the $1/p$-density given in Eq. (\ref{1/r-pot}) is placed in 
Eq. (\ref{bossform}), this yields the acceleration (again $D$ will be $1$ AU)
\bea
&&a_{BP/p}(r) = -[(4GL)~\rho_1] ~ D\int^{100}_{20}{dp } ~\times
\nonumber \\ 
&&
\left[\frac
{\mathbf{K}\left(\sqrt{\frac{4pr}{r^2+2pr+p^2}}\right)}{r(r+p)} 
+\frac{
\mathbf{E}\left(\sqrt{\frac{4pr}{r^2+2pr+p^2}}\right)}{r(r-p)}
\right]  \label{athinr}\\
&\equiv& -[(8GL~\rho_1)/2]~g_{BP/p}(r) \label{bossform/p-1}\\
&=&-(C_1/2) ~g_{BP/p}(r) =-(2/\pi)~C_{ball} ~g_{BP/p}(r).     
    \label{bossform/p-2}
\eea

The numerical integration yielding $g_{BP/p}(r)$ has to deal with integrable singularities at $p=r$, which exist because there the argument of 
$\mathbf{K}$ is unity.   By avoiding the singularities, the integral is doable, except for the two singularities coming from the discontinuous nature of the ring's density at the boundaries.  The result agrees numerically with the result in the previous subsection.  That is, 
\be
a_{BP/p}(r) = a_{T/p}(r), ~~~~~~~~~~~~~
g_{BP/p}(r) = 2~g_{T/p}(r).
\ee


\subsection{\label{boss-peale}
The Boss-Peale model}

Eq. (\ref{bossform}) is the integral used by Boss and Peale \cite{boss} to study gravity from a smooth cylindrical mass distribution of the form
\bea
&& \rho_{BP}(p) = \frac{\rho_0^{BP} (p-A)^2}{D^2}
                 \exp\left[-\frac{(p-A)}{5}\right], \\
&& A = 50 ~ \mathrm{AU} \le p \le 100 ~ \mathrm{AU} = B, 
        ~~D= 1~ \mathrm{AU}. 
\eea
For comparison we take this model to have the same mass, ${\cal M}_{ring}$, given in Eq. (\ref{mring}).  Therefore, 
\bea
{\cal M}_{ring} &=& 4\pi D~\rho^{BP}_0~D^2 \int^{100}_{50} 
             dp~p(p-50)^2 ~\times \nonumber \\
&~~~~~~&
\exp\left[-\frac{(p-A)}{5}\right] \nonumber \\
&=&4\pi D~\rho^{BP}_0~D^2~5^4~[25.8826]. 
\eea
(If one makes the approximation that the upper limit of the integral goes to infinity, then the last term in the second line would be 
$26 =[\Gamma(4)+10~\Gamma(3)]$.)  Therefore, 
\be
\rho^{BP}_0 = \frac{64}{(25.8826)\cdot 25}~\rho_1
= (0.172)\times 10^{-16}~\mathrm{g/cm}^3.
\ee

If we place this density in Eq. (\ref{bossform}) we obtain 
\bea
&& a_{BP}(r) = -C_{BP}~g_{BP}(r), \\
C_{BP} &=& (4GD\rho^{BP}_0)=\frac{8}{5^3\cdot (25.8826)}C_1 \nonumber \\
         &=& 0.002473~C_1 = (0.00687) \times 10^{-8}~\mathrm{cm/s}^2, 
\eea
where the quantity $g_{BP}(r)$ is 
\bea
g_{BP}(r) = \int^{100}_{50}
{dp~p}~(p-50)^2 ~\exp\left[-\frac{(p-50)}{5}\right] ~\times
\nonumber \\
\left[\frac
{\mathbf{K}\left(\sqrt{\frac{4pr}{r^2+2pr+p^2}}\right)}
{r(r+p)} 
+\frac{
\mathbf{E}\left(\sqrt{\frac{4pr}{r^2+2pr+p^2}}\right)}
{r(r-p)}\right].
\eea

\begin{figure}[h]
 \begin{center}
\noindent    
\psfig{figure=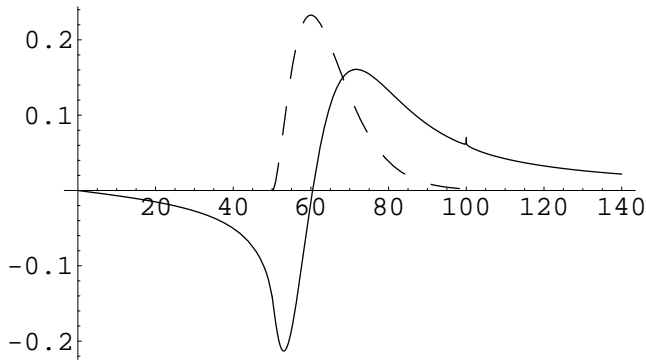,width=3.375in}
\end{center}
  \caption{A plot (solid line) of $-a_{BP}(r)$ in units of $10^{-8}$ cm/s$^2$ 
vs. $r$ in AU.  Also shown is  a dashed plot of $\rho_{BP}(p)$ 
in units of $10^{-15}$ g/cm$^3$.
 \label{bossBP}}
\end{figure}

As in the last subsection, $g_{BP}(r)$ can be integrated numerically \cite{boss}, but with difficulty because of the integrable singularities when $r=p$.  The result for $-a_{BP}(r)$ is shown in Figure \ref{bossBP}, which agrees with Figure 1 of Ref. \cite{boss} (except for the small, narrow spike at $B=100$ -- see below).


Particular values of the acceleration are 
\bea
&-&a_{BP}(\{10,53,73,120\}\mathrm{AU}) = \nonumber \\
      ~~~&&     \{-0.00686,-0.212,+0.159,+0.0325\} 10^{-8}~\mathrm{cm/s}^2.
\nonumber \\ &~&
\label{a-BP/p}
\eea
These values, and the shape of Figure \ref{bossBP} reflect the different type of density profile of this ring.  
Note that the curve for $-a_{BP}(r)$ is smooth when $r=50$.  This is because the density varies continuously from zero at this point.  On the other hand, note the small, narrow spike at $r=100$, which occurs since the "thin" ring abruptly ends there with the density $\rho_{BP}(p)/\rho_{BP}^0$ going discontinuously from $(2500 \exp[-10]) = 0.113$ to zero.  If the ring density is allowed to smoothly continue on past $r=100$, decreasing exponentially out to infinity, the spike disappears and the resulting $-a_{BP_\infty}(r)$ becomes very slightly higher (lower) in magnitude than $-a_{BP}(r)$ going somewhat further out (in) from the position of the spike. 

A comparison of the normalized acceleration, $a_{BP}(r)$, with that for other models will be given in Section \ref{discuss}


\section{\label{3-d-1-r} 
3-d, cylindrical-coordinate, $\mathbf{(1/p)}$-density ring}

\begin{figure}[h]
 \begin{center}
\noindent    
\psfig{figure=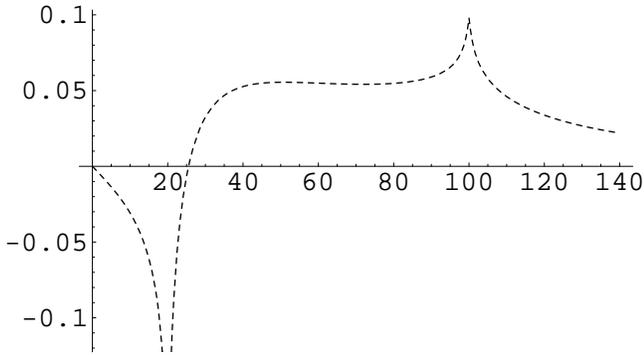,width=3.375in}
\end{center}
  \caption{ A plot of $-a_{1/p}(r)$ in units of $10^{-8}$ cm/s$^2$ vs. distance in AU, obtained from a $(1/p)$-density ring, with a 3-dimensional calculation.  
 \label{cylfig}}
\end{figure}

Now we calculate the acceleration from the ${(1/p)}$-density in the 3-d case. Begin with the complete, exact, 3-dimensional integral defined in Eqs. (\ref{3-dV}) and (\ref{3-da}) with the 
ring ${(1/p)}$-density of Eq. (\ref{1/r-pot}):
\begin{widetext}
\be
a_{1/p}(r) =-G\left[\frac{-d}{dr_x}\right]   
\int^{2\pi}_0 d\phi 
\int^{+D}_{-D} dz 
\int^{R_2}_{R_1}\frac{dp~p~\rho_1L}
{p~\sqrt{p^2 +r_x^2 - 2rp\cos\phi +(z-r_z)^2}}.  \label{2go} 
\ee
Going to the plane of the ecliptic, performing the $p$-integration (which is easy since the density cancels the volume element), and then doing the $r$-derivative yields
\bea
a_{1/p}(r)&\equiv&- (C_1/4)~g_{1/p}(r)  \\
a_{1/p}(r)&=&-\frac{C_1}{4}\left[\frac{-d}{dr}\right] \int^{\pi}_0 d\phi 
\int^{+D}_{-D}dz 
~\times~\ln\left[p - r \cos\phi + \sqrt{p^2 + r^2 + z^2 - 2 p r \cos\phi}\right]^{R_2}_{R_1}   \nonumber \\
&=&-\frac{C_1}{4} \int^{\pi}_0 d\phi \int^{+D}_{-D}dz ~
[\Phi(r,R_2,z,\phi) - \Phi(r,R_1,z,\phi)], \\
\Phi(r,R,z,\phi)&=&\frac{\cos\phi + (R \cos\phi - r)/S}
               {(R - r\cos\phi) + S}         \nonumber \\
&=& \left[\frac 
{-r\sin^2\phi}{z^2+ r^2\sin^2\phi}
\right] +\left[\frac
{S\cos\phi +[-(R^2+r^2)\cos\phi +pR (1 +\cos^2\phi)]/S}
{z^2+ r^2\sin^2\phi}
\right], \label{curly}\\
S&\equiv& \sqrt{R^2 +r^2 - 2rR\cos\phi +z^2} .
\eea
The $z$-integration can be done analytically using the two sets of square brackets in Eq. (\ref{curly}) separately,  with the complicated second piece adding an additional part to the first term.  This  yields
\bea
H(r,R,Z,\phi)&=& -2~ \sin\phi~
\tan^{-1}\left[\frac{Z}{r \sin\phi}\right] + \cos\phi~\ln[Z + S] 
 + \sin\phi~\tan^{-1}\left[\frac{rZ~\sin\phi}
{R^2 + r^2 - 2 R r\cos\phi + (-R + r \cos\phi) S}\right], \nonumber \\
&& \\
S&{\rightarrow}&
        \sqrt{R^2 +r^2 - 2rR\cos\phi +Z^2}.
\eea

Although it is technically possible to do the $\phi$-integration, the result is so complicated that it is preferable to do the final integral numerically.  The result, 
\be
g_{1/p}(r) = \int^{\pi}_0 d\phi~ \left[
H(r,R_2,D,\phi) - H(r,R_2,-D,\phi) -H(r,R_1,D,\phi)+H(r,R_1,-D,\phi)\right],
\label{final-p}
\ee
\end{widetext}
is used to obtain $a_{1/p}(r)$, which is shown in Figure \ref{cylfig}.  (The numerical singularities to be overcome occur when $(r-R_{\{1,2\}}\cos\phi)=0$. 


This figure again has the correct behaviour.  It is a more delicate version of the ``thin ring" shown in Fig. \ref{2-dKB}.  
The most noticeable change from the ``thin ring" is that the spikes of $-a_{T/p}(r)$ near $\{R_1,R_2\}$ in Figure \ref{2-dKB} become finite cusps at $\{R_1,R_2\}$ of $-a_{1/p}(r)$ in Figure \ref{cylfig}.  The cusps are also less extreme compared to the spikes.  This is because, for the 3-d ring, all the near-by mass is {\it not} at a point on the ecliptic, but along a line perpendicular to it.  The proper limit can be seen by evaluating both the 2-d and 3-d forms as $r$ becomes large.  By $r=1000$ the two forms already agree to three significant figures.  

To summarize:  A $1/p$-density potential in a ring does {\it not} produce a constant acceleration within the ring. 

A comparison of the normalized acceleration, $a_{1/p}(r)$, with that for other models will be given in Section \ref{discuss}.  


\section{\label{constring} Cartesian, constant-density ring}

We next consider a constant density disk.  This is of interest for both physical and mathematical comparisons.  We use cartesian coordinates because for cartesian coordinates the volume element is unity.  Therefore, a constant density has the simplest integrals with these coordinates. 
(We already observed how the $1/\sqrt{x^2 + y^2}$ density cancels the 
$\sqrt{x^2 + y^2}$ volume element in cylindrical coordinates.) This current calculation is similar to that used in Ref. \cite{cylinder} 
to study the metrology of a solid cylinder for big-G Cavendish experiments.  

To settle on $\rho_0$ we take the same total mass and shape as the $1/p$ ring.  This means 
\be 
\rho_0 = \rho_1 \frac{2L}{R_1+R_2} = \rho_1/3.
\ee
Now proceed by using Eq. (\ref{3-dV}), giving
\begin{widetext}
\be
{\cal V}_{Con}(r)=-G\rho_0\left[\int^{R_2}_{-R_2}dy 
                \int^{\sqrt{R_2^2 - y^2}}_{-\sqrt{R_2^2 - y^2}}dx
-  \int^{R_1}_{-R_1}dy 
\int^{\sqrt{R_1^2 - y^2}}_{-\sqrt{R_1^2 - y^2}}dx\right] 
    \int^{+D}_{-D}
       \frac{dz}{\sqrt{(x-r_x)^2 + y^2 + (z-r_z)^2} }. 
\ee

The two integrals represent the gravitational effect of a disk of radius $R_2$ minus the effect of a disk of radius $R_1$, thus yielding a ring.   
Again, in the plane of the ecliptic ($r_z=0$) the acceleration is obtained by taking the negative of the derivative of the integrand with respect to $r$:
\be  
-\frac{d}{dr}\frac{1}{[(r-x)^2 + y^2 + z^2]^{1/2}} 
= \frac{r-x}{[(r-x)^2 + y^2 + z^2]^{3/2}}.  
\ee
(Note that since one is taking the derivative of the square root of a square, one must be careful that the correct over-all sign emerges.)   

Now do the integral with respect to $z$.  This yields  
\be
I_z = \left[\frac{z(r-x)}
{[(r-x)^2 + y^2]~[z^2 + (r-x)^2 + y^2]^{1/2}}\right]^D_{-D}.
\ee
Thus, 
\be
a_{Con}(r) =-G\rho_0 
\left[\int^{R_2}_{-R_2}dy\int^{\sqrt{R_2^2 - y^2}}_{-\sqrt{R_2^2 - y^2}}dx
- \int^{R_1}_{-R_1}dy\int^{\sqrt{R_1^2 - y^2}}_{-\sqrt{R_1^2 - y^2}}dx\right]  \frac{2D(r-x)}{[(r-x)^2 + y^2]~[D^2 + (r-x)^2 + y^2]^{1/2}}.
\label{3-dVcart}
\ee
The $x$ integral is 
\be 
I_x =
\ln\left[+D +\sqrt{D^2 + (r-x)^2 + y^2}\right]^{R_2}_{R_1} 
   - \ln\left[-D +\sqrt{D^2 + (r-x)^2 + y^2}\right]^{R_2}_{R_1}, 
\label{x-const}
\ee
so
\bea
a_{Con}(r) &=&-G\rho_0 
\left[\int^{R_2}_{-R_2}dy~F(r,y,R_2,D)
- \int^{R_1}_{-R_1}dy~F(r,y,R_1,D)\right],  \\
F(r,y,R,D)&=& 
\ln\left\{\left[\frac{[+D +\sqrt{D^2 + R^2 + r^2 -2r\sqrt{R^2 - y^2}}]}
{[-D +\sqrt{D^2 + R^2 + r^2 -2r\sqrt{R^2 - y^2}}]}\right] 
\left[\frac{[-D +\sqrt{D^2 + R^2 + r^2 +2r\sqrt{R^2 - y^2}}]}
{[+D +\sqrt{D^2 + R^2 + r^2 +2r\sqrt{R^2 - y^2}}]}\right]\right\}. \label{xInt}
\eea
\end{widetext}

This final integral can be done analytically using involved transformations similar to those used in Ref. \cite{cylinder}. But the end result is very complicated.
Therefore, for clarity, a simple 1-dimensional numerical integral will be used.  
(As a result we leave unaddressed the implications of the relative sizes of 
$r$ vs. $\{R_1,R_2\}$, which implications can play in the analytic form of this final integral.) 
We change all units to AU, e.g., change the variable $y$ to $t=y/D$ and multiply the external constants by the same $D=1$ AU.  Then,
\bea
a_{Con} &=& -C_0 \left[\int^{100}_{-100}dt ~F(r,t,100,1)\right. \nonumber \\
 &~~~~&~~~~~~~~ - \left. \int^{20}_{-20}dt ~F(r,t,20,1)\right] \\
 &=& -C_0 ~ g_{Con}(r), \\
C_0&=& G\rho_0D = C_1/480 = 0.00579~\times 10^{-8} ~\mathrm{cm/s}^2.
\nonumber \\ &~&
\eea

\begin{figure}[h]
 \begin{center}
\noindent    
\psfig{figure=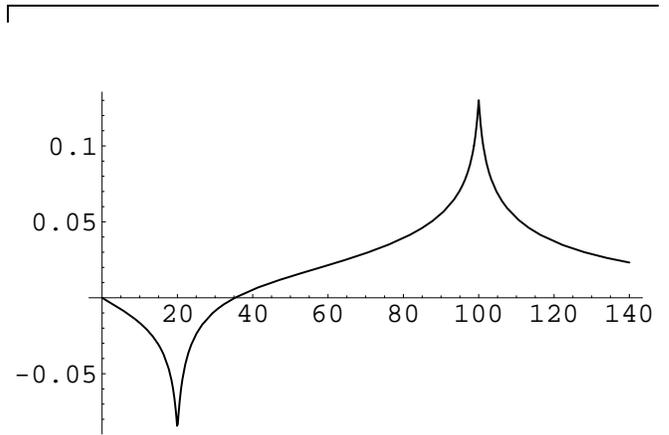,width=3.375in}
\end{center}
  \caption{A plot of $-a_{Con}(r)$ in units of $10^{-8}$ cm/s$^2$ vs. distance in AU for a uniform ring.
 \label{cylgrav}}
\end{figure}

In Figure \ref{cylgrav} we show $-a_{Con}(r)$.  Again we see the correct general behaviour.  With the 3-d calculation, the cusps at the discontinuous boundaries of the ring are large, but finite and hence physical.  Interesting values of the acceleration are 
\bea
-a_{Con}(\{10,\sim 20,60,\sim 100,~120\}~\mathrm{AU})\times 10^{8}~= &&
       \nonumber \\
  \{-0.0165,-0.0870,+0.03146,+0.130,+0.371\}~\mathrm{cm/s}^2. &&
\nonumber \\ &&
\label{a-Con}
\eea
Since the total mass is the same as for the $1/p$ ring, the acceleration should tend to the same limit as $r$ gets large, and it does.  


A comparison of the normalized acceleration, $a_{Con}(r)$, with that for other models will be given in Section \ref{discuss}.


\section{\label{sphwedge}
Wedge $\mathbf{1/r^2}$ (thin-ring $\mathbf{1/p^2}$) density}


\subsection{Wedge configuration}

Now we consider a wedge-shaped slice with the spherical density
\be
p_2(r) = \frac{\rho_2~L^2}{r^2}.
\ee
As before, the slice goes between $R_1$ and $R_2$, except in spherical distance from the origin.  The opening wedge angle is
\be
\theta_0 = \tan^{-1}(D/R_1) = 0.049958 ~\mathrm{radians}.
\ee
Keeping the mass of the slice the same, 
\bea
{\cal M}_{ring} &=& 2\pi (2\delta)\rho_2L^2(R_2-R_1), \\
\delta &=& \sin \theta_0 = 1/\sqrt{401} = 0.049938,
\eea
one has, 
\be
\rho_2 = D/(\delta~L)~\rho_1 \equiv \beta~\rho_1= (1.0012)~\rho_1.
\ee

In the plane of the ecliptic the acceleration from the wedge is 
\bea
a_{1/r^2}(r) =-G\left[\frac{-d}{dr}\right]   
\int^{\pi/2 +\theta_0}_{\pi/2 -\theta_0}d\theta~\sin\theta
\int^{2\pi}_0 d\phi ~\times &&\nonumber \\
\int^{R_2}_{R_1}
\frac{\rho_2L^2}{t^2}
\frac{t^2~dt}{\sqrt{t^2 +r^2 - 2rt~\cos\phi~\sin\theta }}.&&  
\label{wgo} 
\eea
Because the density-functional again cancels the volume element, the $t$-integral yields 
\be 
{\cal I}_t = \ln\left[t - r \sin\theta~\cos\phi + \sqrt{t^2 + r^2 - 2 t r \sin\phi~\cos\theta}\right]^{R_2}_{R_1}.
\ee
Now taking the negative of the $r$-derivative yields
\bea
a_{1/r^2}(r) =-G\rho_2L^2 
\int^{\pi/2 +\theta_0}_{\pi/2 -\theta_0}d\theta~\times 
~~~~~~~~~~~~~~~&&
\nonumber \\  
\sin\theta
\int^{2\pi}_0 d\phi~ U(r,R_1,R_2,\theta,\phi), \\
U(r,R_1,R_2,\theta,\phi) = ~~~~~~~~~~~~~~~~~~~~~~~~~~~~~~~~~~&& 
\nonumber \\
\left[
\frac{(-\sin\theta~\cos\phi)S_t + r-t\sin\theta~\cos\phi}
{[t-r\sin\theta~\cos\phi + S_t]S_t}\right]^{R_2}_{R_1}, &&\\
S_t = \sqrt{t^2 +r^2 - 2rt~\cos\phi~\sin\theta }. 
~~~~~~~~~~~~~~~~&&
\eea

The $\phi$-integral is completely analytic, and yields 
\bea
{\cal I}_\phi(r,R_1,R_2,\theta) &=& \left[
\frac{4t}{rS_-}~\mathbf{K}\left(\sqrt{\frac{-4tr~\sin\theta}{S_-^2}}\right)
\right]^{R_2}_{R_1},  \label{r2k}\\
&=&  \left[
\frac{4t}{rS_+}~\mathbf{K}\left(\sqrt{\frac{4tr~\sin\theta}{S_+^2}}\right)
\right]^{R_2}_{R_1},  \label{r2kk} \\
S_\pm &\equiv& \sqrt{t^2 +r^2 \pm 2rt~\sin\theta }.
\eea

\begin{figure}[h]
 \begin{center}
\noindent    
\psfig{figure=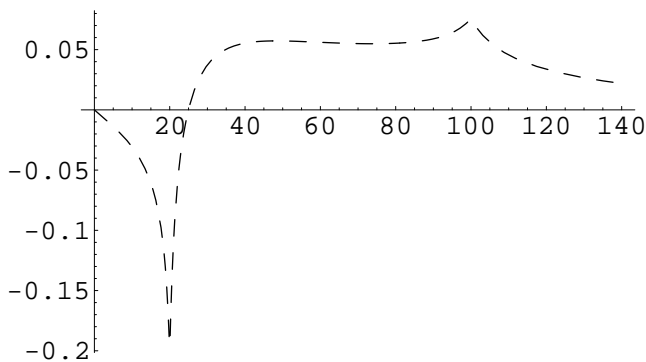,width=3.375in}
\end{center}
  \caption{A plot of $-a_{1/r^2}(r)$ in units of $10^{-8}$ cm/s$^2$ vs. $r$ in AU. 
 \label{wedgefig}}
\end{figure}

We thus have 
\bea
a_{1/r^2}(r) &=&-C_2~g_{1/r^2}(r) \\
C_2&=& G\rho_2L^2/D = C_1/(8 \delta)=2.5031 ~ C_1 \\
g_{1/r^2}(r) &=& D\int^{\pi/2 +\theta_0}_{\pi/2 -\theta_0}d\theta~\sin\theta~
{\cal I}_\phi(r,R_1,R_2,\theta).
\eea
This integral can be done numerically and is used in $a_{1/r^2}(r)$, shown in Figure \ref{wedgefig}.  (The only numerical singularity problems are if both $\theta = \pi/2$ and also $r$ is either $R_2$ or $R_1$.)  


The  most interesting observation is that this result is very similar to that from the $1/p$-density cylindrical ring shown in Figure \ref{cylfig}.  (This point will be shown even better in Section \ref{discuss}.) The fact that the density is falling off faster with distance ($1/r^2$ vs. $1/p$) is compensated for by the increasing spherical width, which is growing as $r~\sin\theta_0$.  \\

A comparison of the normalized acceleration, $a_{1/r^2}(r)$, with that for other models will be given in Section \ref{discuss}.


\subsection{Thin-ring configuration}

To demonstrate the correctness of the assertion that the growing width of the wedge with distance caused the wedge to behave more like a $1/p$ ring, we now quickly look at the ``thin-ring"  $1/p^2$ problem.  Keeping the same mass as before and using the formalism of Section \ref{numTR} yields (also see Eq. (\ref{athinr}))
\bea
\rho_{T/p^2}(p) &=& \frac{\rho_{T2} L^2}{p^2}, \\
\rho_{T2}&=& \frac{(R_2-R_1)}{L \ln(R_2/R_1)} = (2.485)~\rho_1\\
a_{T/p^2}(r) &=&  -C_{T2}~g_{T/p^2}(r),  \\
 C_{T2} &=& C_1\frac{(R_2-R_1)}{4D~\ln(R_2/R_1)} = (24.85)~C_1, \\
g_{T/p^2}(r) &=&  D^2\int^{R_2}_{R_1} dp~
\left[\frac
{\mathbf{K}\left(\sqrt{\frac{4pr}{r^2+2pr+p^2}}\right)}{p~r(r+p)} \right.
\nonumber \\
&&~~~~~~ +~\left.\frac{
\mathbf{E}\left(\sqrt{\frac{4pr}{r^2+2pr+p^2}}\right)}{p~r(r-p)}
\right]  \label{athinr2}.
\eea

\begin{figure}[h]
 \begin{center}
\noindent    
\psfig{figure=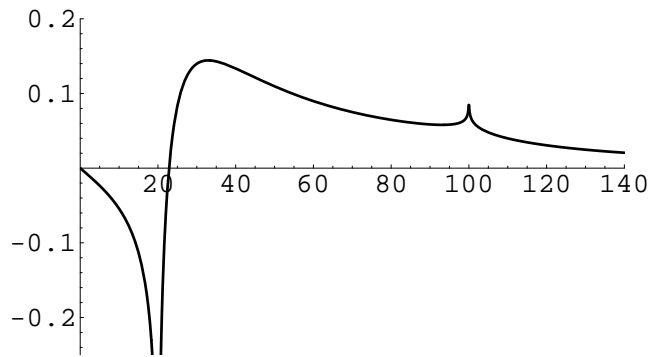,width=3.375in}
\end{center}
  \caption{A plot of $-a_{T/p^2}(r)$ in units of $10^{-8}$ cm/s$^2$ vs. $r$ in AU.
 \label{thin2}}
\end{figure}

In Figure \ref{thin2} we show a plot of $a_{T/p^2}(r)$. 
One clearly sees the difference between the $1/r^2$ wedge and the $1/p^2$ thin ring.  With its rise going inward within the ring, $-a_{T/p^2}(r)$ displays the higher mass concentration at $r=R_1$.  (Again there is the thin-ring caveat that the spikes at $r =\{R_1,R_2\}$ would be finite cusps in a 3-d calculation.)  


A comparison of the normalized acceleration, $a_{T/p^2}(r)$, with that for other models is also given in Section \ref{discuss}.


\section{Discussion} 
\label{discuss}

The different physical models we have investigated in this paper provide an intuitive understanding about what type of accelerations can be obtained from Kuiper Belt models.  In particular, they can not easily yield a constant (or even an approximately constant) gravitational acceleration in a cylindrical system.  

As to the specific gravitational accelerations in the plane of the ecliptic, $a(r)$, we found:  

\begin{itemize} 
\item
Starting out with Figure \ref{gshell}, one sees that even a spherical shell with a $1/r$ density only yields an approximately constant acceleration near the outer edge of the shell.  
\item
Continuing on to a ``thin ring" with sharp edges, the $1/p$ density produces an acceleration that is singular at the edges of the ring and is approximately constant near the middle of the ring. (See Figure \ref{2-dKB}.)  
\item
Contrary to this, the smoother-density, ``thin-ring" Boss-Peale  model produces a smooth acceleration at the inner edge and shows only a slight, narrow spike if the density has a small discontinuous jump at the outer edge instead of decreasing smoothly to infinity.  (See Figure \ref{bossBP}.)  Thus, the physical differences in shape between the $1/p$ and Boss-Peale models end up being instructive.
\item
The full 3-dimensional, $1/p$ model, yields a finite acceleration everywhere, so the cusps at the edges of the ring are finite compared to the spikes of the 2-dimensional ``thin-ring" approximation.  (See Figure \ref{cylfig}.)
\item
The 3-dimensional constant-density ring produces softer cusps yet a more undulatory variation than the $1/p$ ring.  (See Figure \ref{cylgrav}.)  It is intermediate in its effects between the 3-d, $1/r$ ring and the 2-d, Boss-Peale ``thin" ring.
\item
The 3-dimensional wedge, with a spherical fall off in density of $1/r^2$, produces an acceleration that is very similar in shape to that from the $1/p$ cylindrical ring.  (See Figure \ref{wedgefig}.) This is because the growing width of the wedge with distance approximately makes up for the faster fall off of density with distance.
\item
The above assertion is demonstrated by the contrasting behaviour of the $1/p^2$-density, thin ring's $a_{T/p^2}(r)$, compared to the wedge's $a_{1/r^2}(r)$.  It varies much more in the belt and reaches a high maximum near $r=R_1$.  (See Figure \ref{thin2}).
\end{itemize} 
(We also mentioned how to extend these results to out of the plane of the ecliptic by taking $r_z \neq 0$ and then studying both $a(r_x)$ and $a(r_z) = -(d/dr_z){\cal V}(\mathbf{r})$.)


\begin{figure}[h]
 \begin{center}
\noindent    
\psfig{figure=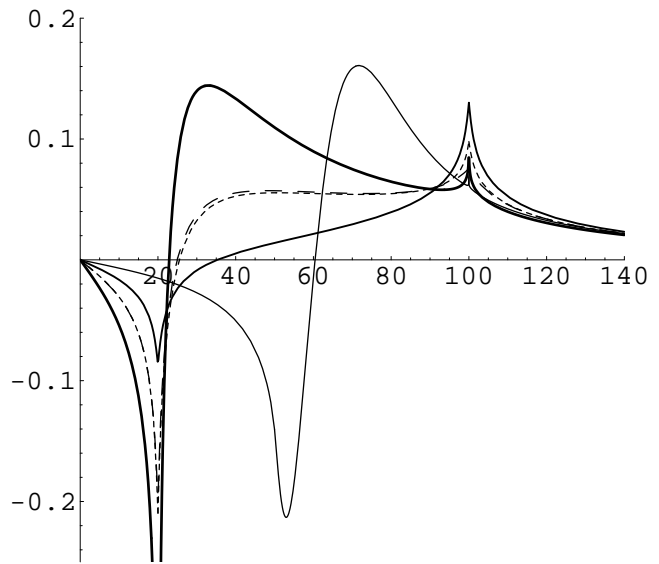,width=3.375in}
\end{center}
  \caption{Plots of  $[-a(r)]$, in units of $10^{-8}$ cm/s$^2$, vs. distance 
(in AU), 
for: (i) the 3-d, $1/p$-density ring, $[-a_{1/p}(r)]$ -- short-dashed line, (ii) the Boss-Peale 2-d ``thin ring," $[-a_{BP}(r)]$-- narrow line, (iii) 
the constant-density ring, $[-a_{Con}(r)]$ -- medium line, (iv) the $1/r^2$-density, wedge, $[-a_{1/r^2}(r)]$ -- dashed line, and (v) the $1/p^2$-density, "thin ring," $[-a_{T/p^2}(r)]$ -- wide line. 
 \label{aAll}}
\end{figure} 


The results emphasize how difficult it is to achieve a truly constant acceleration within a finite cylindrically-symmetric system (not even considering how much mass would be needed to mimic the Pioneer anomaly).  
This difficulty can be put in mathematical context.  
Consider just the ``thin ring," which is mathematically simpler than the full 3-d ring.  Starting with Eq. (\ref{2-dV}), a constant acceleration between $R_1$ and $R_2$ would be produced by a density $\rho_C(p)$ that satisfied 
\be
r = \mathrm{Const.}~\int^{R_2}_{R_1}{dp~p}~\rho_C(p)~I_\phi(r),
\ee
where $I_\phi(r)$ is given in Eqs. (\ref{I1}) and (\ref{I2}). That is a  complicated inverse problem.  Formally it could be done by a decomposition into cylindrical harmonics, but that is not the point here.

Finally, in Figure \ref{aAll}, we show a direct comparison of the physical accelerations of (i) the
3-d, $1/p$ ring, (ii) the 2-d, Boss-Peale ``thin ring," (iii) the 3-d, constant ring, (iv) the 3-d, $1/r^2$ wedge, and (v) the "thin," $1/p^2$ ring,  all with the same total 
mass, 1.96 $M_\oplus$.   (As before we cut off the infinite spikes at the boundaries of the thin rings.)   When $r\rightarrow \infty$, all the curves tend to $[G{\cal M}_{ring}/r^2]$, as they should.  This is even though the differing density distributions produce quite different accelerations within the ring. 

To within normalizations, the results in Figure \ref{aAll} agree with the type of results published previously for Kuiper-Belt disks \cite{boss,pioprd}.  Most importantly, within the ring the acceleration is not constant.  Further, especially in the central portions of the rings, the accelerations are approximately two orders of magnitude too small to explain the Pioneer anomaly.


\begin{acknowledgments}

I thank Claus L\"ammerzahl and J. D. Anderson 
for their valuable comments, criticisms, and suggestions.  
The support  of the U.S. DOE is acknowledged.     

\end{acknowledgments}



\end{document}